# Emergence of superconductivity near 11 K by suppressing the 3-fold helical-chain structure in noncentrosymmetric HgS

He Zhang[1,2], Wei Zhong[3], Yanghao Meng[1,2], Bowen Tang[1], Binbin Yue[3,*], Xiaohui Yu[1,2,4,*], and Fang Hong[1,2,4,*]

[1]Beijing National Laboratory for Condensed Matter Physics, Institute of Physics, Chinese Academy of Sciences, Beijing 100190, China

[2]School of Physical Sciences, University of Chinese Academy of Sciences, Beijing 100049, China

[3]Center for High Pressure Science and Technology Advanced Research, 10 East Xibeiwang Road, Haidian, Beijing 100094, China

[4]Songshan Lake Materials Laboratory, Dongguan, Guangdong 523808, China

**ABSTRACT:** The trigonal α-HgS has a 3-fold helical chain structure, and is in form of a noncentrosymmetric $P3_121$ phase, known as the cinnabar phase. However, under pressure, the helical chains gradually approach and connect with each other, finally reconstructing into a centrosymmetric NaCl structure at 21 GPa. Superconductivity emerges just after this helical-nonhelical structural transition. The maximum critical temperature ($T_c$) reaches 11 K at 25.4 GPa, $T_c$ decreases with further compression, and is still 3.5 K at 44.8 GPa. Furthermore, the $T_c$-critical magnetic field ($B_{c2}$) relation exhibits multi-band features, with a $B_{c2}$ of 5.65 T at 0 K by two-band fitting. Raman spectra analysis demonstrates that phonon softening plays a key role in structural transition and the emergence of superconductivity. It is noted that HgS is the first reported IIB group metal sulfide superconductor and the only NaCl-type metal sulfide superconductor with a $T_c$ above 10 K. This work will inspire the exploration of superconductivity in other chiral systems and will extend our understanding of the versatile behavior in such kinds of materials.

**INTRODUCTION**

Binary metal sulfides constitute a vast material family. Numerous metal sulfides are essential natural minerals and crucial in scientific study. For example, $MoS_2$, $WS_2$, and other layered sulfides have attracted people's interest due to their outstanding electrical and optical properties[1-3]; CdS-based nanomaterials are promising carriers in photocatalytic hydrogen production[4]; $PbS$[5-7] and $HgS$[8] quantum dots have great potential in photoelectric device applications. Binary metal sulfides show diverse forms and functions. They can be generally classified into several different classes according to stoichiometry: transition metal tri-chalcogenides (TMTCs), transition metal dichalcogenides (TMDCs), and metal mono-sulfides (MMSs) as well as other stoichiometry sulfides like sesquisulfides. Typical TMTCs exhibit quasi-1-dimensional (1D) properties, TMDCs are commonly 2D layered hexagonal structures, and MMSs are generally in form of 3D structures.

The properties of these sulfides can be manipulated in many ways like atomic intercalation[9] and substitution[10].

The modulated properties can be expanded exponentially based on the large family of sulfides. Pressure, a fundamental thermodynamic parameter, could modify material properties efficiently without changing the composition or adding impurities by controlling the atom distance and electron overlap in a wide range. In particular, layered TMDCs are sensitive to the external pressure. Pressure can help to realize insulator-metal transition and superconductivity (SC) in various layered binary metal sulfides[11-13]. At the same time, it is also an important tool to study the correlation between SC and charge density wave (CDW) in $NbS_2$, and $TaS_2$[14-16]. Therefore, by applying pressure, binary metal sulfides will show more abundant phenomena and they are also ideal platforms for studying pressure-induced phase transition and superconductivity.

Different from weak interlayer interactions existing in 1D TMTCs and 2D TMDCs, the 3D MMSs have more compact atomic interactions and seem to show much higher structural stability under pressure. However, previous studies demonstrate that 3D metal mono-sulfides have complex phase transition routes and distinguished properties under pressure. Here, we summarize the structural phase transition evolution sequence and superconductivity information (transition pressure and $T_c$) of MMSs under pressure in **Table 1,** according to the group number: 1) Except for BeS, all alkaline earth sulfides[17-20] have a NaCl structure at ambient pressure and go through a NaCl–CsCl phase transition upon compression. In the same group, when the metal cation mass increases, the cation/anion radius ratio rises, so the cations prefer to select a configuration with a higher coordination number (from NaCl-type with coordination number 6 to CsCl-type with coordination number 8) at lower pressure. While BeS[21] has the zinc-blende (ZB) structure under ambient pressure and turns into the $P6_3/mmc$ (NiAs) structure under pressure. 2) IIIB rare earth metal sulfides generally have a NaCl structure with superconductivity at ambient pressure. IIIB metal Sc, Y, and La sulfides undergo a phase transition from NaCl structure to CsCl structure just like IIA sulfides, and the critical pressure decreases from ScS to LaS. 3) The phase transition evolution of IVA tin[22] and lead sulfides[23, 24] are similar, starting with the non-SC structure at ambient pressure and progressing via an orthorhombic intermediate phase to the superconducting CsCl structure. GeS has the *Pnma* structure at ambient pressure and via the *γ-Pnma* structure turns into the *Cmcm* structure[25], higher pressure phase transition is calculated[26] while no experiment is reported. 4) As for transition metals from IVB-VIII group elements, like V, Cr, Mn, etc., with unfilled d electrons, show various oxidation states and complex sulfide structures. Many transition metal mono-sulfides have the NiAs-type structure at ambient pressure, like VS, CrS, FeS, CoS, and NiS except for MnS. Besides the NiAs structure, VS has the *Pnma* structure and NiS has the R3m structure. FeS has many polymorphous structures at different temperatures and pressure, such as tetragonal ($P4/nmm$), monoclinic ($P2_1/c$), *Pnma*, $P6_3mc$, *P-62c*, NiAs ($P6_3/mmc$). Superconductivity is observed in tetragonal FeS[27, 28], with a $T_c$ of 4.1 K at ambient pressure and suppressed by pressure. Then a structural transition to a hexagonal (*P-62c*) phase happens at 7.2 GPa, with another superconductivity dome, and a maximum $T_c$ of 6 K is found at 16.1 GPa. For MnS, there are NaCl, ZB, and wurtzite ($P6_3mc$) structures at ambient pressure, called α-MnS, β-MnS, and γ-MnS respectively. β-MnS, γ-MnS would turn into α-MnS under pressure of 5.3 and 2.9 GPa[29]. The vacancy in NaCl-type ZrS is an important variable tuning the

**Table 1:** The structure evolution and superconductivity of common binary metal mono-sulfides under pressure.

| Group | Sulfide | Structure | | | Superconductivity | | |
|---|---|---|---|---|---|---|---|
| | | Low pressure phase | $P_c$/GPa | High pressure phase | SC structure | $T_c^{max}$/K | $P$/GPa |
| IIA | BeS[21] | (ZB) | 59 | (NiAs) | - | - | - |
| | MgS[17] | (NaCl) | 158 | (CsCl) | - | - | - |
| | CaS[17] | (NaCl) | 40 | (CsCl) | - | - | - |
| | SrS[19] | (NaCl) | 18 | (CsCl) | - | - | - |
| | BaS[20] | (NaCl) | 6.5 | (CsCl) | - | - | - |
| IIIB | ScS*[30] | (NaCl)* | 87.2 | (CsCl) | (NaCl) | 5.1 | 0 |
| | YS*[31] | (NaCl)* | 49 | (CsCl) | (NaCl) | 2.8 | 0 |
| | LaS*[32] | (NaCl)* | 16.8 | (CsCl) | (NaCl) | 0.84 | 0 |
| IVB - VIII | ZrS*[33] | (NaCl)* | | | (NaCl) | 2.4 | 0 |
| | | (WC)* | | | (WC) | 3.7 | 0 |
| | VS | (NiAs) | | | - | - | - |
| | | *Pnma* | | | - | - | - |
| | CrS | (NiAs) | | | - | - | - |
| | | (NaCl) | | | - | - | - |
| | MnS[29] | (ZB) | 5.3 | (NaCl) | - | - | - |
| | | (Wurtzite) | 2.9 | (NaCl) | - | - | - |
| | FeS*[27, 28] | *P4/nmm** | 7.2 | *P-62c** | *P4/nmm* | 4.1 | 0 |
| | | | | | *P-62c* | 6 | 16.1 |
| | | polymorphous | | | - | - | - |
| | CoS | (NiAs) | | | - | - | - |
| | NiS[34] | (NiAs) | | Not clear | - | - | - |
| | | *R3m* | | | - | - | - |
| | PdS*[35] | *P42/m* | 19.6 | Not clear* | Not clear | 6 | 24.6 |
| IB | CuS*[36, 37] | (NiAs) | 15 | amorphous | | | |
| | | *Cmcm** | 15 | amorphous | *Cmcm* | 1.62 | 0 |
| IIB | ZnS[38] | (Wurtzite)/(ZB) | 12.2 | (NaCl) | - | - | - |
| | | (NaCl) | 65 | *Cmcm* | | | |
| | | *Cmcm* | 215 | (CsCl) | | | |
| | CdS[39] | (Wurtzite) | 4.3 | (NaCl) | - | - | - |
| | | (NaCl) | 52 | *Cmcm* | | | |
| | | *Cmcm* | 70 | *Pmmn* | | | |
| | | *Pmmn* | 129 | distorted (CsCl) | | | |
| | | distorted (CsCl) | 361 | (CsCl) | | | |
| | <span style="color:red">HgS*(This work)</span> | <span style="color:red">Cinnabar</span> | <span style="color:red">21</span> | <span style="color:red">(NaCl)*</span> | <span style="color:red">(NaCl)</span> | <span style="color:red">11</span> | <span style="color:red">25.4</span> |
| IVA | GeS[25, 40] | *Pnma* | 15 | *γ-Pnma* | - | - | - |
| | | *γ-Pnma* | 30 | *Cmcm* | | | |
| | | Cmcm | 72 | (NaCl) | | | |
| | SnS*[22] | *Pnma* | 8.1 | *Cmcm* | (CsCl) | 5.8 | 47.8 |
| | | *Cmcm* | 18 | (CsCl)* | | | |
| | PbS*[23, 24] | (NaCl) | 4.5 | *Cmcm* | (CsCl) | 12.3 | 19.1 |
| | | *Cmcm* | 22 | (CsCl)* | | | |

Critical pressure ($P_c$), maximum superconducting critical temperature ($T_c^{max}$), and corresponding pressure ($P$) are listed. Non-SC transition metal sulfides are exemplified by the 4th-period sulfides. Asterisk (*) marks SC compounds and phases. 0 pressure means the sample is superconducting at ambient pressure. Grey shading highlights IIA and IIB group sulfides.

superconductivity[41], besides NaCl-type, WC (*P-6m*2) structure ZrS is also superconducting[33]. PdS has a semiconducting tetragonal structure at ambient pressure and shows metallic behavior and superconductivity under pressure of 29 GPa, with a $T_c$ of 6 K. The high pressure structure of PdS is not reported[35]. 5) CuS, an IB sulfide, undergoes a structural transition from the NiAs ($P6_3/mmc$) structure to the *Cmcm* structure below 55 K, with a superconducting $T_c$ of ~1.62 K[37]. With increasing pressure, CuS turns into an amorphous phase at ~15 GPa[36]. 6) In IIB sulfides, ZnS with ZB or wurtzite structure turns into NaCl structure at 12.2 GPa, further through an intermediate orthorhombic phase *Cmcm*[38] to CsCl structure. A similar phase transition happens in CdS[39], from wurtzite to NaCl and eventually to CsCl, with another extra intermediate phase *Pmmn*.

Among the IIA+IIB (II) group metal sulfides highlighted in gray in **Table 1**, we notice a blank area in superconductivity research. Will some of these sulfides such as HgS or ZnS become superconductors under pressure? This is a question worth investigating. HgS is a good prototype of II group MMSs. Unlike insulating IIA MMSs, HgS has 5d valance electrons, which may show a sensitive electronic behavior under high pressure. There are two different crystal forms of HgS: one is the α-HgS or cinnabar, and it is an insulator in red color; the other one, β-HgS or metacinnabar, is a gray semimetal with a ZB structure meta-stable phase. More interesting, α-HgS has a unique noncentrosymmetric trigonal lattice, while S and Hg atoms alternately interconnect to form 3-fold helical chains parallel to the *c* axis. The space group (S.G.) of the cinnabar phase can be expressed as $P3_121$ (S.G.152) or $P3_221$ (S.G.154), and both of them have a 3-fold helical chain structure but with different helix directions. The chiral crystal structure leads to asymmetry properties. Chiral phonons splitting of single crystal α-HgS was observed by circularly polarized Raman spectroscopy[42]. α-HgS stands out from other metal mono-sulfides due to its unique helical chain structure, providing a motivation to investigate the pressure effect on the helical structure like modulating the bandgap, electronic structures, and inducing helicity transition.

In this work, we reported the superconductivity in α-HgS by suppressing the helical chain structure under pressure and provided a clear picture to show the correlation between structure and electronic behavior. Consistent results were obtained from the electrical transport, X-ray diffraction, second harmonic generation (SHG) measurements, Raman spectroscopy, absorption and reflection spectroscopy. As pressure increases, the insulating α-HgS first undergoes a direct-indirect bandgap transition near 8 GPa, followed by a structural phase transition from the helical chain structure to the cubic structure near 21 GPa confirmed by XRD and SHG experiments, accompanied by the emergence of superconductivity. $T_c$ reaches the maximum of 11 K at 25.4 GPa and then decreases with further compression. Clear phonon softening is observed, which is expected to be responsible for the structure instability of the helical phase and superconductivity. It is noted that HgS is the first superconductor among the II group sulfide family, and it exhibits a relatively high $T_c$, which is also the highest one among the NaCl-type metal sulfide superconductors. This work provides new insights into the mechanics of NaCl-type superconductors and will inspire the exploration of superconductivity in other metal sulfides and chiral systems as well.

**RESULTS AND DISCUSSION**

**Fig. 1a** shows the atomic structures of the cinnabar phase at low pressure and the NaCl phase at high pressure, as well as the schematic diagrams of the 3-fold screw/rotation structures of two phases and their projection onto the plane perpendicular to the 3-fold axis, confirmed by *in situ* high pressure X-ray diffraction. HgS in the cinnabar phase has a 3-fold helical chain structure with a spacing of *a* between the two closest chains and a repeating unit length of *c*. There are three Hg atoms located at ($u$, 0, 1/3), (0, $u$, 2/3), (1-$u$,1-$u$,0) and three S atoms located at ($v$, 0, 5/6), (0, $v$, 1/6), (1-$v$, 1-$v$, 1/2) in the unit cell. Upon compression, the helical chains in HgS get closer. The *a*-axis direction is more compressible than the *c*-axis direction, $u$ and $v$ are getting closer to 1/3. This makes the structure more similar to the NaCl structure. Eventually, the helical structure vanishes when $u=v=1/3$, and the chains are close enough to connect with each other and construct a cubic lattice.

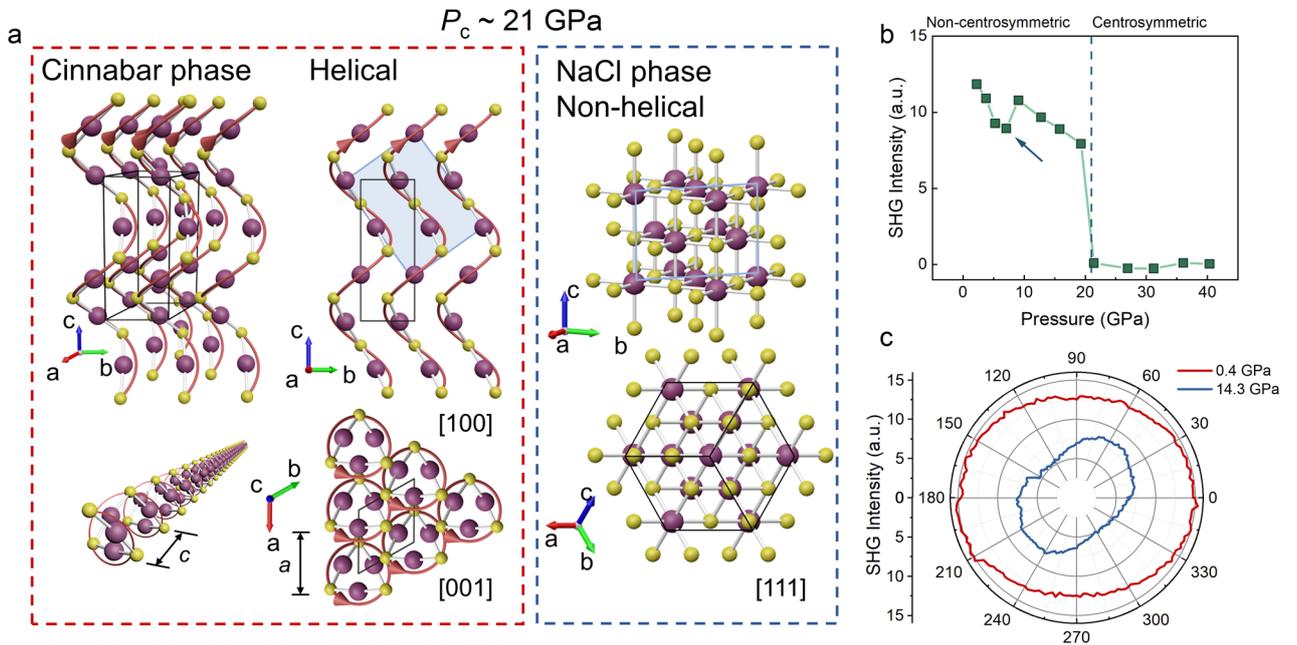

**Fig. 1:** Structure schematic diagrams and SHG measurements of HgS under pressure. **a,** The atomic structures of cinnabar phase $P3_121$ (S.G. 152) at ambient pressure and NaCl phase $Fm$-$3m$ (S.G. 225) at high pressure. Yellow and purple balls represent S atoms and Hg atoms. **b,** Pressure-dependent SHG intensities of HgS, SHG signal disappears at 21 GPa. **c,** Polar plot of SHG signal of HgS under different pressures.

**Table 2:** Structures of HgS at low and high pressures.

| Structure | S.G. | Symmetry | $P$/GPa | $a$/Å | $b$/Å | $c$/Å | | Atom position | Multiplicity |
|---|---|---|---|---|---|---|---|---|---|
| Cinnabar | $P3_121$ | Non-centrosymmetric | 0.3 | 4.136 | 4.136 | 9.472 | Hg | (0,0.281,2/3) | 3 |
| | | | | | | | S | (0,0.501,1/6) | |
| NaCl | $Fm$-$3m$ | Centrosymmetric | 30.1 | 5.118 | 5.118 | 5.118 | Hg | (0,0,0) | 4 |
| | | | | | | | S | (0,0,1/2) | |

Lattice parameters and atom positions of the cinnabar phase and NaCl phase are shown.

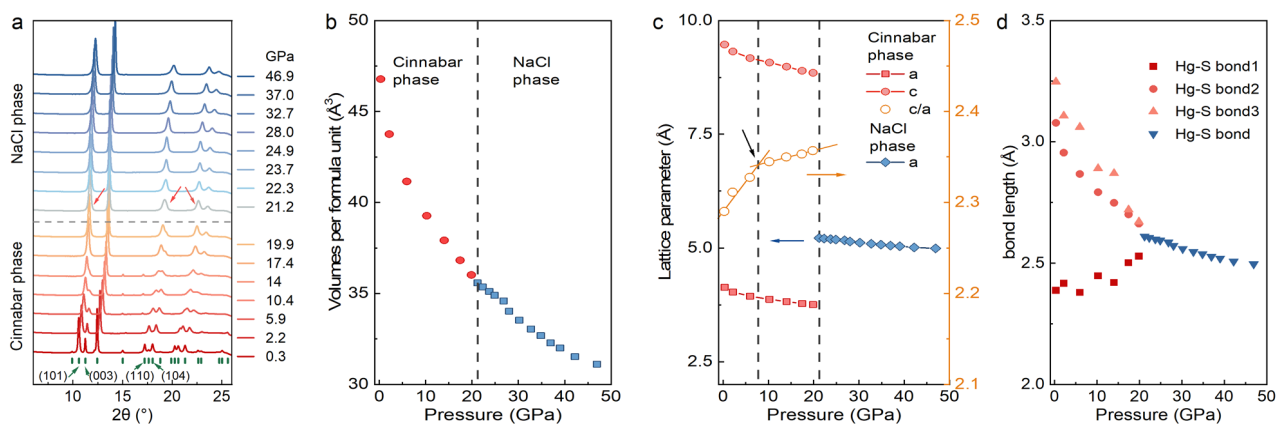

**Fig. 2**: XRD patterns and refined results of HgS under pressure. **a,** The XRD patterns of HgS with the X-ray wavelength of 0.6199 Å under pressure up to 46.9 GPa. Green lines mark the Bragg positions at ambient pressure. **b,** The volume per formula unit of HgS under pressure. **c,** The lattice parameters of the cinnabar phase and NaCl phase under pressure. **d,** Adjacent Hg-S bond lengths of HgS obtained from refined XRD patterns under pressure.

The SHG measurement was carried out under pressure. At low pressures, the non-centrosymmetric structure of HgS is revealed through an apparent SHG signal. As pressure increases, the SHG signal disappears abruptly at ~21 GPa (**Fig. 1b**), marking the non-centrosymmetric to centrosymmetric structural transition, namely cinnabar-NaCl phase transition. The polarization-dependent SHG measurement of powder HgS under pressure was also carried out. As seen in **Fig. 1c**, a slight polarization dependence indicates the grains in the sample HgS. The anisotropy of the polar plot is enhanced by pressure and the maximum intensity position takes a 45° move, suggesting the re-orientation of the grains under pressure

The X-ray diffraction patterns of HgS under pressure are shown in **Fig. 2**. With increasing pressure, the (003) peak at 11.3° weakens and becomes closer to the (101) peak at 10.6°, eventually merging with it. The (104) peak at 18.0° eventually merges with the (110) peak at 17.2° at around 21.2 GPa, suggesting an increased symmetry. There is only one structural transition up to 46.9 GPa, from the cinnabar phase to the NaCl phase around 21 GPa. The diffraction peaks move towards a higher degree with increasing pressure, indicating that the lattice shrinks under compression. The peaks hardly move above 21 GPa, indicating that the cinnabar phase is more compressible than the NaCl phase. This is further confirmed by the equation of states fitting shown in **Fig. S4**, the bulk modulus of the NaCl phase is ~92 GPa, significantly higher than the cinnabar phase, around 32 GPa.

Rietveld refinement[43] is an efficient way to obtain more structural details from the X-ray diffraction patterns. We used GSAS-II refinement software[44] to obtain accurate structural information of HgS under pressure, like lattice parameters, unit cell volumes, and Hg-S bond lengths. To compare the volume of a unit cell in two phases, the volume is divided by multiplicity to get the volume per formula unit. As shown in **Fig. 2b**, the volume changes continuously with pressure before and after the phase transition. Since the space group of the cinnabar phase is the subgroup of that of the NaCl phase[45], we confirmed that the structural transition of HgS from the cinnabar phase to the NaCl phase is a second-order phase transition. The lattice parameters are shown in **Fig. 2c**. Lattice parameter $a$ is more compressible than *c*, with *c/a* increasing with pressure. Detailed information about lattice parameters and atomic positions of the two phases is listed in **Table 2**. By refining atom positions we obtained the Hg-S bond lengths in **Fig. 2d**. There are three types of Hg-S bonds with different bond lengths in the cinnabar phase, Hg-S bond 1 is the bond length on the Hg-S chains, and Hg-S bond 2 and 3 stand for closest Hg-S atom distance of adjacent chains. With increasing pressure, the difference between them becomes smaller, and there is only one Hg-S bond length in the NaCl phase. The XRD and SHG results show the process of the vanishing of helicity in α-HgS under pressure, and the helicity transition is accompanied by many other properties like insulator-metal transition and superconductivity which we will discuss later.

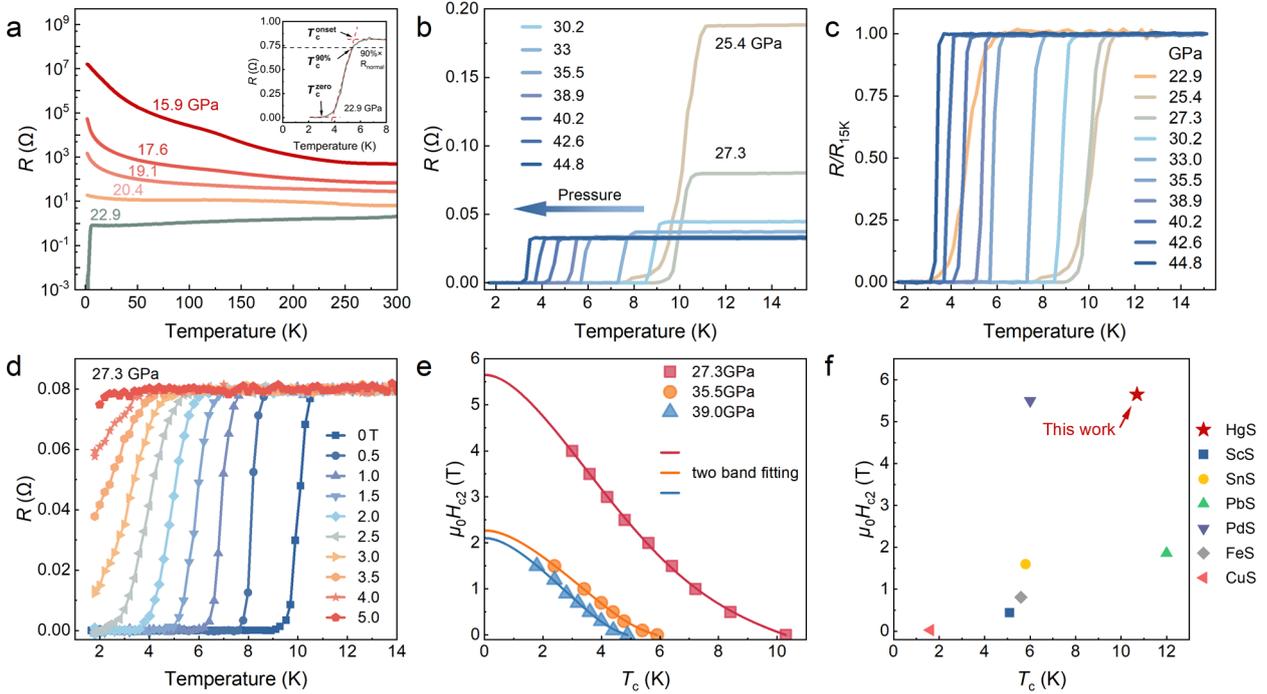

**Fig. 3:** The electrical transport properties of HgS under pressure. **a,** The *R-T* curves from 15.9 to 22.9 GPa. Below 15.9 GPa, the resistance is too large and exceeds the detection limit of the experimental setup. The inset shows the criteria of $T_c^{onset}$, $T_c^{90\%}$, and $T_c^{zero}$. **b,** The *R-T* curves from 25.4 to 44.8 GPa. **c,** Normalized *R-T* curves of HgS near the superconductivity transition. **d,** Superconductivity transition under different external magnetic fields at 27.3 GPa. **e,** Two band fitting of upper critical magnetic fields and $T_c$ relations at three different pressures. **f,** $T_c$ versus upper critical magnetic fields of HgS and other superconducting binary metal mono-sulfides.

**Fig. 3** displays the electrical transport properties of HgS under high pressure. At low pressure < 15.9 GPa, the resistance of the HgS is too large, which is out of the detection limit in our experimental setup. At 15.9 GPa, the sample still behaves as a semiconductor and the resistance increases significantly from ~$10^3$ Ω at 300 K to ~$10^7$ Ω below 2 K, as seen in **Fig. 3a**. The semiconducting behavior persists until 20.4 GPa while the resistance of the sample decreases overall monotonously with pressure, indicating a continuously reduced band gap under pressure. At 22.9 GPa, the sample turns into a metal and shows superconductivity. The $T_c^{onset}$ is ~5.6 K and the zero-resistance state is presented below 3.1 K, as shown in the inset of **Fig. 3a**. Upon further compression, the SC transition becomes shaper,

$T_c$ is enhanced and reaches 11.0 K at 25.4 GPa, above which the superconductivity is gradually suppressed though the metallic behavior of the sample becomes better, as seen in **Fig. 3b**. At 44.8 GPa, $T_c$ decreases to ~3.5 K. To better demonstrate the SC evolution trend, **Fig. 3c** shows the normalized $R$-$T$ curves near the SC transition region. For reference, the electrical transport behavior is also investigated during the decompression process, and it gives a consistent $T_c$-$P$ relationship and critical transition point, showing a phase transition without hysteresis, as seen in **Fig. S1**.

We investigated the magnetic field effect on the superconductivity of HgS up to 5.0 T. At 27.3 GPa, the zero-field $T_c$ is ~10.3 K, and it decreases with increasing field, as seen in **Fig. 3d**. When the magnetic field reaches 5 T, the SC transition occurs below the lowest temperature 1.7 K, and is barely identified. Meanwhile, the SC transition becomes more and more broadening by increasing the external magnetic field. Both phenomena further confirm the superconductivity of HgS. Magnetic field versus $T_c$ under three different pressures is shown in **Fig. 3e**. Here, the $T_c$ is defined by the temperature at which the resistance drops to 90% of the normal state value called $T_c^{90\%}$ ($T_c$ without special instructions stands for $T_c^{90\%}$). The curvature behavior in the low magnetic field range is noteworthy, and the Ginzburg-Landau (GL) fitting is not applicable. Since NaCl-type HgS has a high symmetry structure, the anisotropic single-band superconductivity is excluded, and the curvature at low fields is attributed to multiband superconductivity. The zero-temperature upper critical magnetic field at 27.3 GPa is around 5.65 T by two band fitting. That is much higher than the 1.2 T at 100 GPa for pure sulfur[46, 47]. Previous calculation shows that d electrons of Hg and p electrons of S mainly contribute to the density of states (DOS)[48] in NaCl phase HgS. It is noted that the $T_c$ in pure Hg decreases with pressure (apart from 4.2 K with pressure), while the $T_c$ of sulfur is enhanced with pressure from 10 to 17 K[46, 47]. In this case, p electrons in S may still dominate the contribution to the superconductivity but it is modulated by the crystal field stabilized by Hg. Due to the much larger mass of Hg, the overall phonon frequency will decrease compared with pure S, which results in a declining trend of $T_c$ with pressure in HgS. At 35.5 GPa and 39.0 GPa, due to the decrease of $T_c$, their upper critical fields also decrease, as seen in **Fig. 3e**, and the $R$-$T$ curves at 35.5 GPa and 39.0 GPa are shown in **Fig. S1**. **Fig. 3f** shows the upper critical field and $T_c$ of HgS and other metal mono-sulfides. HgS has the highest upper critical field and high $T_c$ close to 11.0 K.

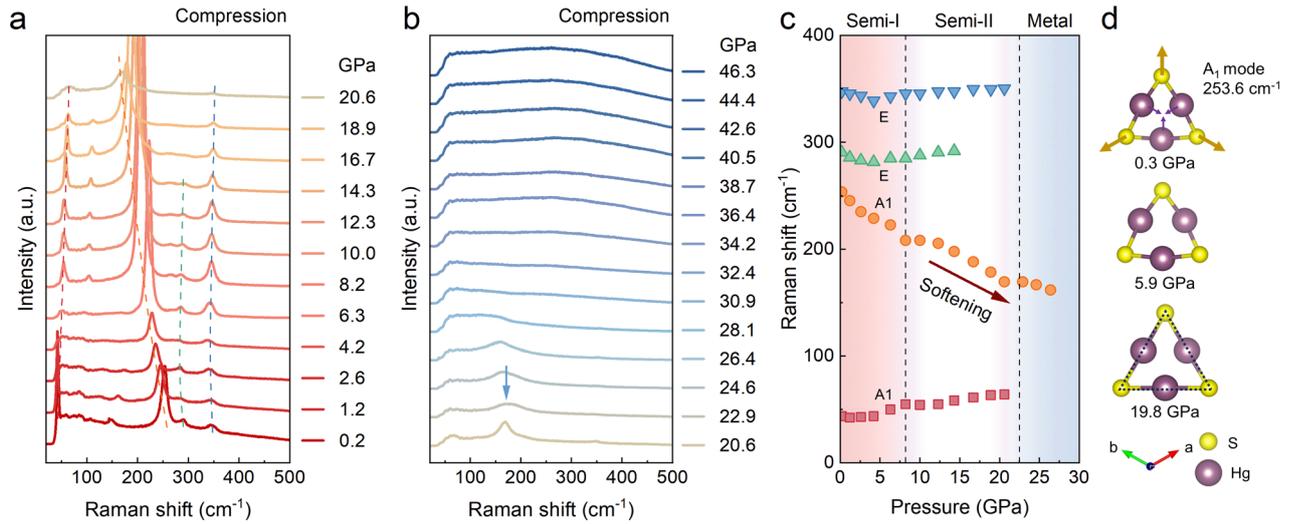

**Fig. 4:** Raman spectroscopy of HgS under pressure. **a-b,** The sequence of Raman spectra of HgS under compression, phase transition, and metallization could be recognized at about 22.9 GPa. **c,** The position of Raman peaks is modulated by pressure. Significant softening of the $A_1$ main peak could be observed. Dash lines separate different phase regions. **d,** The schematic diagrams of atomic motions of 253.6 cm$^{-1}$ $A_1$ mode and the chain-structure evolution projected onto the *ab* plane.

Raman spectroscopy provides valuable insights into meta-excitations in solid materials and is used to investigate the structural transition in HgS under pressure. The cinnabar phase has the point group of $D_3$ and vibrational modes of $2A_1+3A_2+5E$. Two $A_1$ modes are Raman-active, three $A_2$ modes are inferred active, and five E modes are active in both Raman and inferred spectroscopy. The Raman peaks shown in **Fig. 4** are in good agreement with prior ambient pressure studies [49]. The main peaks at 253.6 cm$^{-1}$ and 45.1 cm$^{-1}$ are recognized as $A_1$ modes, called chain breathing modes: atomic displacements in the *ab* plane along the direction perpendicular to the *c* axis. Two frequencies 253.6 cm$^{-1}$ and 45.1 cm$^{-1}$ correspond to two different bases, nearly pure S motions and pure Hg motions[50]. **Fig. 4d** shows the 253.6 cm-1 $A_1$ mode which has the largest intensity. 148.7 cm$^{-1}$, 291.2 cm$^{-1}$, and 347.6 cm$^{-1}$ are recognized as E modes, other weak peaks are hard to recognize. As seen in **Fig. 4a-b**, the main Raman peak at 253.6 cm$^{-1}$ is softened under compression, showing a noticeable shift towards lower frequencies as pressure increases. The softened mode is attributed to the increase of Hg-S bond length of the helical chain, meanwhile, the projection of the helical chain is more likely to be a regular triangle with pressure. The E mode peaks first get softened and then move to higher frequencies. The remaining peaks exhibit a slight shift towards higher frequencies upon compression. The pressure dependence of two $A_1$ modes and some strong intensity E modes is shown in **Fig. 4c**. Based on the discontinuous pressure dependence, three different regions can be seen. The two different semiconductor regions correspond to a direct bandgap below ~8 GPa and an indirect bandgap above ~8 GPa, which will be discussed later based on the results of absorption spectroscopy.

Electron-phonon coupling influences quantum phenomena such as superconductivity. The repulsive Coulomb interaction competes with the attractive interaction provided by electron-phonon coupling, and the latter leads to a conventional superconductivity. BCS theory[51] and McMillan strong-coupled theory[52] point out that electron-phonon coupling λ, Debye temperature Θ (deduced from the phonon Debye frequency), and the repulsive interaction μ* contribute to the $T_c$. A larger λ results in a higher $T_c$. The repulsive interaction is near constant in many materials, so the electron-phonon coupling and Debye frequency play a more important role in determining the $T_c$ of superconductors. Phonon softening of the vibrational modes makes the electron-phonon spectral function distribution shift to a lower frequency, thus enhancing the electron-phonon coupling λ. A large λ is essential to induce phonon-mediated superconductivity. As can be seen in **Fig. 5**, the superconductivity emerges just at the metallization, without a non-SC metal intermediate state. This is also attributed to the softened phonons causing sufficient λ to generate phonon-mediated superconductivity.

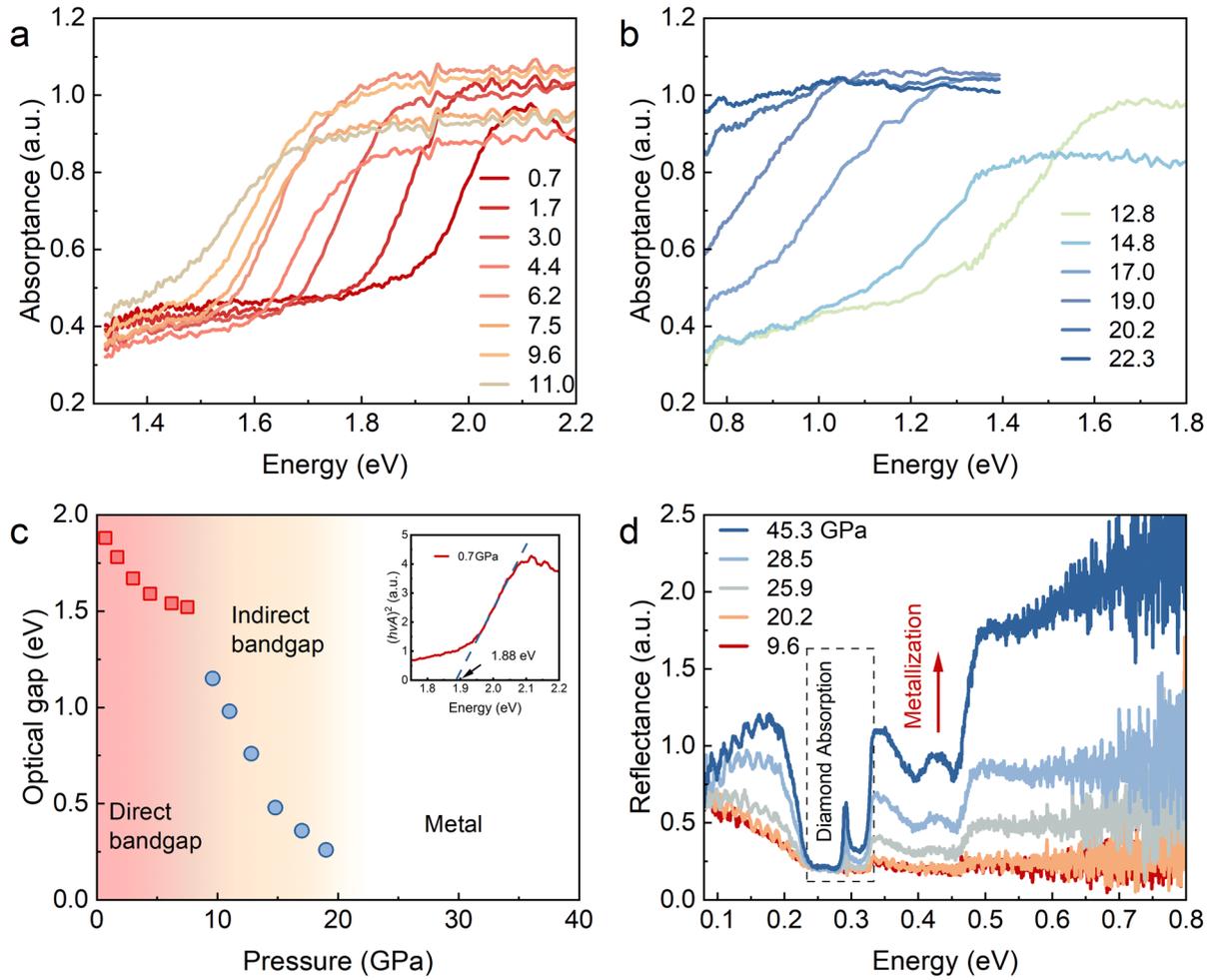

**Fig. 5:** UV-VIS-NIR absorption spectra and middle-inferred (MIR) reflection spectra of HgS under pressure. **a-b,** UV-VIS-NIR absorption spectra. **c,** The optical band gap evolution obtained by the Tauc plot method. The inset shows the linear extrapolation and the optical band gap is the intercept. **d,** MIR reflection spectra show the insulator-metal transition of HgS under pressure.

The Raman peaks exhibit enhanced intensity at 6.3 GPa, and the Raman signals become even more clear than at lower pressures. No structural transition was observed at that pressure. The Raman peaks are gradually suppressed and become very broadening at ~22.9 GPa, which aligns with the structural transition. **Fig. S2** demonstrates that the decompression critical pressure and peak behavior align with those observed during compression, suggesting a phase transition without hysteresis. The abrupt changes in the intensity of Raman peaks at 6 GPa were also observed during decompression, which is associated with an electronic phase transition that we will discuss later, rather than experimental errors. The Raman spectra reveal the critical pressure near 21 GPa for the structural transition, which is consistent with the results from transport measurements and X-ray diffraction.

As mentioned at the beginning, the resistance of the insulating HgS below 15 GPa is too large to be measured by our instrument, hence it is difficult for us to obtain information about the sample's properties below 15 GPa using electrical transport measurements alone. We can learn more about the evolution of electronic structures below 15 GPa by using UV-VIS-NIR and infrared spectroscopies. The UV-VIS-NIR absorption spectra are shown in **Fig. 5a-b**. The absorption edge is clear and moves to lower energy with increasing pressure, which indicates a decreasing band gap. To obtain the optical band gap, the Tauc plot method [53] was used to process the absorption results. At ambient conditions, HgS is a direct band gap insulator. Below 9.6 GPa, the absorption edge red shifts with increasing pressure and approaches to constant. After 9.6 GPa, the absorption gap changes dramatically with pressure, and the absorption edge is slightly widened, an indirect model is preferred in this case. By analyzing the intercept of the linear fitting line, the direct optical band gaps, indirect band gaps, and their pressure dependence are obtained as shown in **Fig. 5c**. Prior first-principle calculations gave the conclusion that the cinnabar phase is a direct band gap semiconductor at pressures below 8 GPa and an indirect band gap semiconductor at pressures above 8 GPa[48]. This matches our findings. **Fig. 5d** shows the middle-inferred (MIR) reflection spectra. It's clear that the reflectance hardly changes below 20 GPa, suggesting an insulating state at a low pressure range, however, above 20.2 GPa, a sudden rise in reflectance was observed, which indicates the metallic transition, which matches our resistance measurements and prior calculations[48].

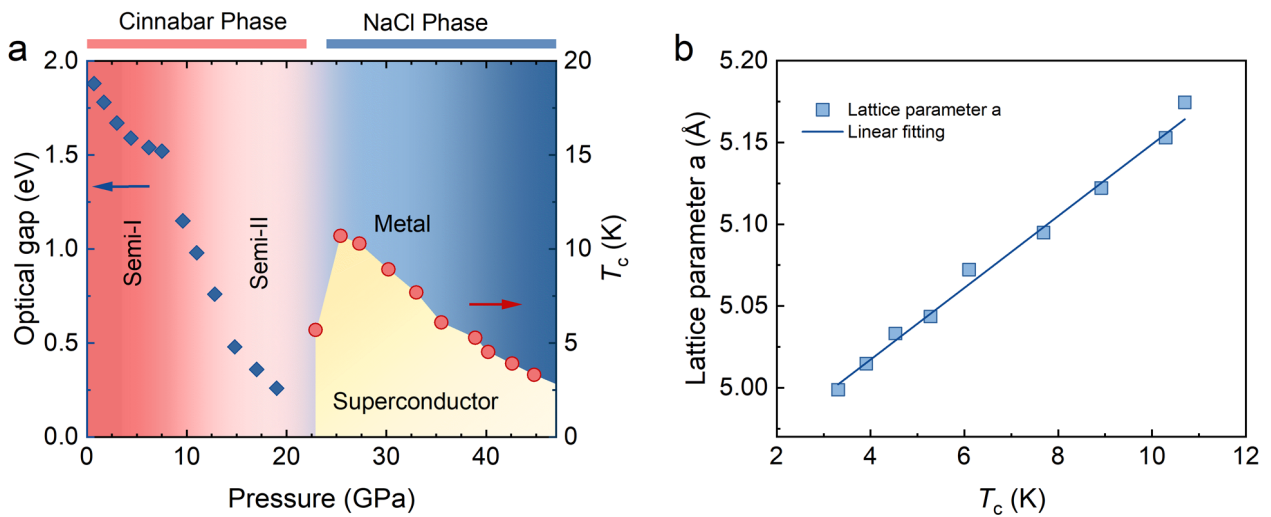

**Fig. 6**: Phase diagram and lattice parameter-$T_c$ relationship of HgS under pressure.

The SHG, X-ray diffraction, electric transport measurements, Raman spectroscopy, and UV-VIS-NIR absorption spectroscopy reveal a distinct structural transition from the cinnabar phase to the NaCl phase at around 21 GPa and show different behavior of those two phases. Based on these results, a phase diagram of HgS under pressure is proposed in **Fig. 6**. Below 21 GPa, cinnabar phase HgS shows typical insulator behavior. Under increasing pressure, the resistance and optical gap of HgS are reduced, and a direct-indirect band gap transition happens at 8 GPa. Then HgS turns into a superconductor with maximum $T_c$ around 11 K above 21 GPa. To study the relationship between lattice parameters and $T_c$, the equation of states fitting is used at each superconducting pressure point. A linear lattice parameter-$T_c$ relationship is observed. Sharing the same structure as NaCl phase HgS, there are other NaCl-type superconductors. These include metal sulfides, metal selenides, metal pnictogenides, and metal carbides[30, 31, 54-56]. For metal sulfides, as previously mentioned, ScS[30] and YS[31] are superconducting at room temperature with a NaCl structure and $T_c$ of 5.1 K and 2.8 K, respectively. NbC and TaC[54] are NaCl-type carbide superconductors with high $T_c$ of 10.8 K and 10.2 K but low upper critical fields of 1.93 T and 0.65 T, respectively. Pnictogenide superconductors SnAs[55], and SnP[56] have $T_c$ of ~3.6 K and 2.8 K. Compared to other NaCl-type superconductors, HgS has relatively high $T_c$ close to 11 K and a high upper critical field of 5.65 T.

We note that trigonal Te and Se also have 3-fold helical chain structures and the same space group $P3_121$ as HgS at ambient pressure[46, 47, 57]. The semiconductor-metal transition of Te happens at 4 GPa, with the structural transition from helical chain structure to monoclinic Te-II phase[58, 59]. Te-II is a superconductor with a strong pressure dependence of $T_c$ around 4 K[59, 60]. Semiconducting Se-I phase undergoes a metal transition to the Se-II phase when the helical structure vanishes, however, no superconductivity occurs. Instead, superconductivity develops in the Se-III phase at 23 GPa, which is isostructural with Te-II [59, 61]. Sulfur adopts a ring molecule structure at ambient pressure. When subjected to pressures > 3 GPa and heated, S can transition into the trigonal chain structure S-II phase, sharing the same structure with trigonal Te and Se. This phase remains stable within a pressure range of 1.5 GPa to 36 GPa at room temperature[62] . Upon further compression, the S-II phase transforms into tetragonal S-III phase at 36 GPa, followed by a metallic phase at 83 GPa[63], eventually demonstrating superconductivity at 100 GPa[46, 47]. In this case, the superconductivity observed in HgS might stem from the chemical pre-compression effect in S by Hg, considering that Hg exhibits a relatively low $T_c$ alone under high pressure.

**CONCLUSION**

We investigated the electric transport properties, X-ray diffraction, SHG measurement, Raman spectroscopy, and UV-VIS-NIR spectroscopy experiments of HgS under pressures up to 45 GPa, and provided a clear phase diagram to show the close correlation between the structure evolution and electronic behavior. XRD and SHG experiments provided a clear structural transition from the noncentrosymmetric cinnabar phase to the centrosymmetric NaCl phase, and the phase transition is accompanied by the emergency of superconductivity. In the superconducting NaCl phase, $T_c$ can reach a maximum close to 11 K, and the SC transition exhibits multi-band features with a $B_{c2}$ of 5.65 T at 27.3 GPa. The phonon softening is very significant and it should be responsible for the structural instability and consequent superconductivity. Notably, HgS is the only NaCl-type metal sulfide superconductor with a $T_c$ exceeding 10 K. Our study on HgS under pressure will stimulate the research of superconductivity in other IIB group metal sulfides and materials with chiral structures.

## EXPERIMENTAL METHODS

Standard four-probe resistance measurements were performed to investigate the electric transport properties of α-HgS under high pressure by using a BeCu-type diamond anvil cell (DAC). The *R-T* curves were collected from 1.7 K to 300 K up to 45 GPa. The *R-T* curves near the superconducting transition region were collected in magnetic fields up to 5.0 T. The DAC has two diamonds with 300 μm culet. A rhenium gasket with a c-BN insulating layer was drilled with a 100 μm hole to serve as a sample chamber. KBr was used as the pressure medium. The HgS powder sample was pre-compacted and cut into a long-strip shape before being loaded into the sample chamber. Four Pt electrodes were contacted to the sample in the standard four-probe configuration. The pressure of the sample chamber was calibrated by the ruby fluorescence at room temperature[64].

A symmetric DAC with Bohler-type diamonds and WC seats was used for *in situ* XRD experiments. A pre-compressed rhenium gasket was drilled with a 120 μm hole to serve as a sample chamber. The powder sample was finely ground and loaded into the sample chamber. Silicon oil was used as the pressure transmitting medium.

Another symmetric DAC with 300 μm culet diamond (Type-II) was used for the SHG experiment, Raman spectroscopy and UV-VIS-NIR spectroscopy under pressure. A 304 stainless steel gasket was pre-compressed and drilled with a 120 μm hole as a sample chamber. KBr was used as the pressure medium. SHG experiment was measured in a home-designed optical system (Ideaoptics, China). The exciting light source is a 1064 nm fiber laser. And SHG signal was collected by a photomultiplier tube (Thorlabs Inc., PMT1000). Raman spectroscopy under pressure was operated by a commercial confocal Raman spectrometer. 532 Nd-YAG laser was used to get the Raman spectra of HgS under pressure. *In situ* high pressure ultraviolet-to-visible-to-near-infrared (UV-VIS-NIR) absorption spectroscopy was performed on the home-designed spectroscopy system (Ideaoptics, Shanghai, China). The high-pressure infrared reflection spectroscopy experiments were performed at room temperature on a Bruker VERTEX 70v infrared spectroscopy system with a HYPERION 2000 microscope.

## ACKNOWLEDGMENTS

This work was supported by the National Key R&D Program of China (Grants No. 2021YFA1400300), the National Natural Science Foundation of China (Grants No. 11921004 and No.11820101003). The *in situ* XRD measurements were performed at 4W2 High Pressure Station, Beijing Synchrotron Radiation Facility (BSRF), which is supported by Chinese Academy of Sciences (Grants No. KJCX2-SW-N20 and No. KJCX2-SW-N03). Part of the experimental work was carried out at high-pressure synergetic measurement station of Synergetic Extreme Condition User Facility (SECUF).